\documentclass[twocolumn,prb,amsmath,amssymb,floatfix,showpacs]{revtex4}
\usepackage{graphicx}
\usepackage{dcolumn}
\usepackage{bm}
\usepackage{times}
\begin{document}


\title{ Magnetic field induced enhancement of spin-order peak intensity in La$_{1.875}$Ba$_{0.125}$CuO$_4$}
\author{Jinsheng Wen}
\author{Zhijun Xu}
\author{Guangyong Xu}
\author{J.~M.~Tranquada}
\author{Genda Gu}
\affiliation{Condensed Matter Physics and Materials Science
Department, Brookhaven National Laboratory, Upton, New York 11973,
USA}
\author{S.~Chang}
\author{H.~J.~Kang}
\altaffiliation{New address: Department of Physics \&\ Astronomy,
Clemson University, Clemson, SC 29634-0978, USA} \affiliation{NIST
Center for Neutron Research, National Institute of Standards and
Technology, Gaithersburg, Maryland 20899, USA}
\date{\today}

\begin{abstract}
We report on neutron-scattering results on the impact of a magnetic
field on stripe order in the cuprate
La$_{1.875}$Ba$_{0.125}$CuO$_4$. It is found that a 7~T magnetic
field applied along the {\it c} axis causes a small but finite
enhancement of the spin-order peak intensity and has no observable
effect on the peak width. Inelastic neutron-scattering measurements
indicate that the low-energy magnetic excitations are not affected
by the field, within experimental error. In particular, the small
energy gap that was recently reported is still present at low
temperature in the applied field. In addition, we find that the
spin-correlation length along the antiferromagnetic stripes is
greater than that perpendicular to them.
\end{abstract}


\maketitle

The role that charge- and spin-stripe orders play in the
superconductivity of cuprates has been quite controversial. It is
commonly believed that the stripe order is harmful for pairing,
given the fact that the superconducting temperature $T_c$ vs hole
content $x$ curve shows an anomaly at $x=1/8$ for
La$_{2-x}$Ba$_x$CuO$_4$, La$_{2-x}$Sr$_x$CuO$_4$, and
La$_{1.6-x}$Nd$_{0.4}$Sr$_x$CuO$_4$, where static spin-stripe order
is observed.~\cite{moodenbaugh,koike1,prb104517,tranquada1} However,
there has been recent evidence from transport and susceptibility
measurements showing that the stripe order is compatible with
pairing and two-dimensional (2D) superconductivity, although it can
inhibit three-dimensional (3D) superconducting phase
order.~\cite{prl67001,johnnew}

One possible way to explore the correlation between
superconductivity and spin-stripe order is to apply a magnetic field
and study the spin order. In La$_{2-x}$Sr$_{x}$CuO$_4$
(Refs.~\onlinecite{nature299,prbr14677,chang-2007}), and
La$_2$CuO$_{4+\delta}$
(Refs.~\onlinecite{PhysRevB.66.014528,PhysRevB.67.054501}), there
are field induced intensity enhancements of the elastic
incommensurate magnetic peaks observed by neutron scattering. The
intensity growth follows the prediction of Demler {\it et
al.},~\cite{prl67202} who analyzed a model of coexisting but
competing phases of superconductivity and spin-density-wave order.
In contrast, it has been reported that the magnetic field has no
impact on the pre-existing stripe order in La$_{2-x}$Ba$_x$CuO$_4$
($x=0.095$)\cite{dunsiger-2007} and
La$_{1.6-x}$Nd$_{0.4}$Sr$_x$CuO$_4$ ($x=0.15$).\cite{prb184419} In
all of these cases, the applied field causes $T_c$ to decrease, but
the onset temperature of the magnetic order remains constant or
increases slightly. Rather surprisingly, a transverse-field muon
spectroscopy study\cite{savici:157001} found a substantial field
induced enhancement of the muon-spin-relaxation ($\mu$SR) rate for
La$_{2-x}$Ba$_x$CuO$_4$ with $x=1/8$, suggesting increases in both
the onset temperature for quasistatic magnetic order and the
low-temperature hyperfine field.

An applied magnetic field can also affect magnetic excitations. For
example, the spin gap observed\cite{LSCO2} in optimally- and
over-doped La$_{2-x}$Sr$_{x}$CuO$_{4}$ is readily modified by an
applied field.~\cite{science1759,LSCO3,chang-2007} In a separate
paper,\cite{johnnew} we report on the observation of a rather small
spin gap of $\sim$0.7~meV at low temperature in
La$_{1.875}$Ba$_{0.125}$CuO$_4$. It would be exciting if this gap
were associated with superconductivity; however, it could also be
due to spin-orbit exchange-anisotropy effects, as for
antiferromagnetic spin waves.\cite{PhysRevB.37.9761} The two
possibilities are potentially distinguishable by testing the impact
of a magnetic field.

To gain insight into the issues discussed above, we carried out
elastic and inelastic neutron-scattering measurements on
La$_{1.875}$Ba$_{0.125}$CuO$_4$ to look at the magnetic field effect
on the spin-stripe order and low-energy magnetic fluctuations. In
this Brief Report, we will show that the main effect of a magnetic
field along the {\it c} axis is to slightly enhance the spin-order
peak intensity, while the peak width and the low-energy magnetic
excitations, as well as the gap feature, remain unchanged (within
experimental uncertainty). By analyzing the spin-order peak width,
we find that the correlation length parallel to the stripes is
larger than that perpendicular to them.

\begin{figure}[ht]
\includegraphics[width=0.9\linewidth]{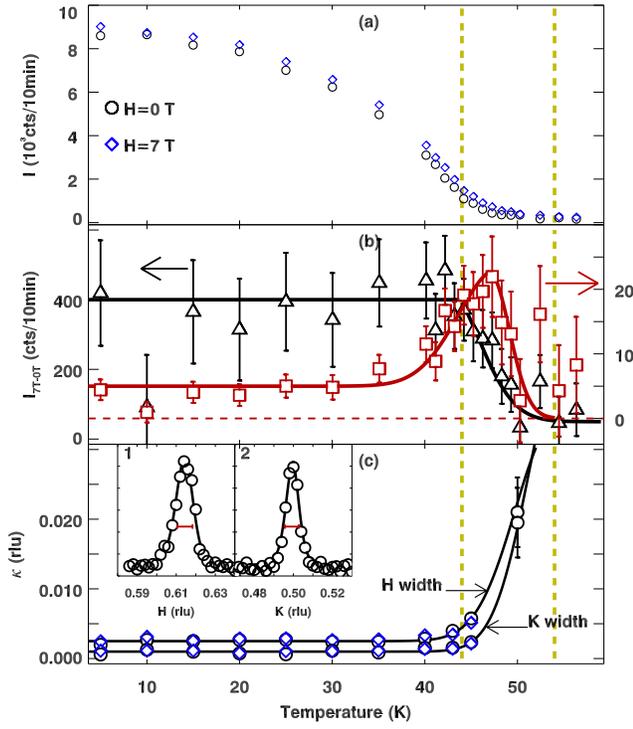}
\caption{(Color online) Spin-order peak (0.615,0.5,0) intensity and
width. (a) Background subtracted peak intensity in zero field
(circles) and 7~T field (diamonds). (b) Peak intensity difference
between 7 and 0~T measurements (triangles), and relative intensity
difference {\it S} defined as $(I_{7 T}-I_{0 T})/I_{0 T}$ (squares).
(c) Resolution corrected peak width along $(0.615+h,0.5,0)$ and
$(0.615,0.5+k,0)$ in zero field (circles) and 7~T field (diamonds).
Insets in (c) show scan profiles along $H$ and $K$ directions. Lines
through the data are guides for the eyes. Vertical lines denote the
onset temperatures, as discussed in the text. Two horizontal lines
in the insets show the instrumental resolutions (FWHM). Error bars
represent $1\pm\sigma$ uncertainties determined assuming Poisson
statistics, and those smaller than the symbols are absent.}
\label{fig:1}
\end{figure}

The single crystal of La$_{1.875}$Ba$_{0.125}$CuO$_4$ used here, a
cylinder with a diameter of 8~mm and a length of 35~mm, was grown in
an infrared image furnace by the floating-zone technique. It is the
same crystal used in Ref.~\onlinecite{johnnew}, with bulk $T_c$ of
$\sim$5~K, and 2D superconducting correlations appearing at the
temperature $T_c^{2D}\sim$40~K. Neutron-scattering experiments were
carried out on the triple-axis spectrometer SPINS located at the
NIST Center for Neutron Research using beam collimations of
$55'$--$80'$--$S$--$80'$--open ($S=$sample) with fixed final energy
of 5~meV. The (002) Bragg reflection from highly-oriented pyrolytic
graphite crystals was used to monochromatize the incident and
scattered neutrons. A cooled Be filter was put after the sample to
reduce contamination from higher-order reflections of the analyzer.
All data were taken in the ($HK$0) scattering plane defined by the
vectors [100] and [010] in tetragonal notation and described in
terms of reciprocal lattice unit (rlu), where 1~rlu~$=a^*=2\pi/a=
1.661$~\AA$^{-1}$. With the sample mounted in a vertical-field
superconducting magnet, the applied field was parallel to the $c$
axis of the crystal.

In Fig.~\ref{fig:1} we plot the background subtracted spin-order
peak intensity (obtained by sitting at the peak position and
counting) and width (obtained by fitting scans through the peak) as
functions of temperatures in zero field and in a field of 7~T. In
zero field, the peak intensity starts to grow at $\sim$54~K, higher
than the temperature, $\sim$42~K, where the peak width reaches its
minimum value. The situation here is similar to that in
La$_{1.6-x}$Nd$_{0.4}$Sr$_x$CuO$_4$, where the nominally elastic
signal detected at higher temperature was attributed to integrated
intensity of low-energy spin fluctuations.\cite{PhysRevB.59.14712}

After cooling in a 7~T magnetic field, there is small but clear peak
intensity enhancement, as shown in Fig.~\ref{fig:1}(a). However, the
peak width, either along $H$ or $K$, is not noticeably affected.
When we plot the difference between $H=7$~T and $H=0$~T measurements
[Fig.~\ref{fig:1}(b)], it can be seen that the difference grows as
the spin order develops, with the same onset temperature as the
zero-field peak intensity, and reaches a maximum when the peak width
saturates. When taking into account the relative intensity
difference $S$, defined as $(I_{7 T}-I_{0 T})/I_{0 T}$, one can see
that it reaches a maximum near 46~K, just before the zero-field
onset of static spin ordering.  This behavior suggests a slight
increase in the spin-ordering temperature, a result qualitatively
consistent with the $\mu$SR results.\cite{savici:157001}

When looking at the peak width [see Fig.~\ref{fig:1}(c)], we found
that the width for the scan along ${\bf Q}=(0.615+h,0.5,0)$ is
larger than that for the scan along $(0.615,0.5+k,0)$. Those widths
are obtained by fitting the data with a Lorentzian function
convolved with Gaussian function representing the instrumental
resolution. The resolutions [full width at half maximum (FWHM)] at
(0.615,0.5,0) along $H$ and $K$ directions are 0.0078 and
0.0072~rlu, respectively. Insets 1 and 2 in Fig.~\ref{fig:1}(c) show
scan profiles along $H$ and $K$ directions at 5~K, from which one
can see that the $H$ scan FWHM is slightly above resolution FWHM,
while the $K$ scan is almost resolution limited. From these scans,
it appears that the correlation length parallel to the
antiferromagnetic stripes is greater than that perpendicular to
them.

\begin{figure}[ht]
\includegraphics[height=0.95\linewidth,angle=90]{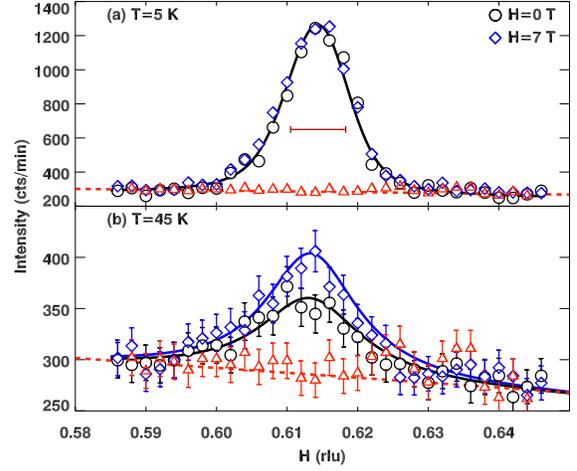}
\caption{(Color online) Selected elastic scans along ${\bf Q} =
(H,0.5,0)$ in zero field (open circles) and 7~T field (diamonds) at
5 and 45~K. Solid lines are guides to the eye. The triangles show
55~K data as the background, as indicated by the dashed lines. The
horizontal line in (a) shows the instrumental resolution (FWHM).
Error bars represent the square root of the counts, and those
smaller than the symbols are absent.} \label{fig:2}
\end{figure}

\begin{figure}[ht]
\includegraphics[height=\linewidth,angle=90]{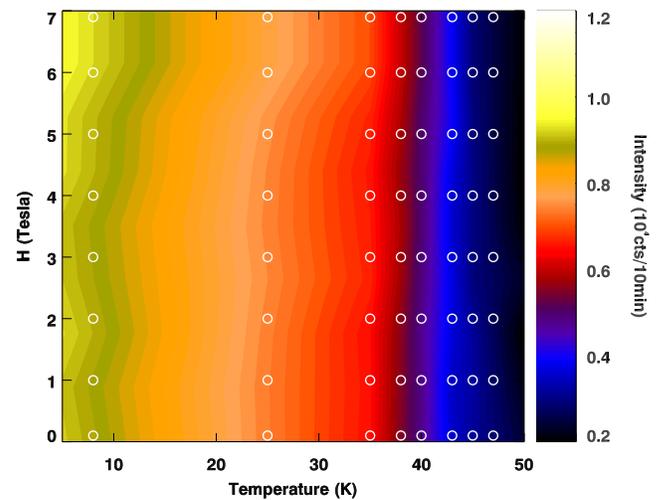}
\caption{(Color online) Contour map of the spin-order peak
(0.615,0.5,0) intensity as a function of temperature and magnetic
field. Circles indicate the fields and temperatures at which the
measurements were performed.} \label{fig:3}
\end{figure}

\begin{figure}[ht]
\includegraphics[width=\linewidth]{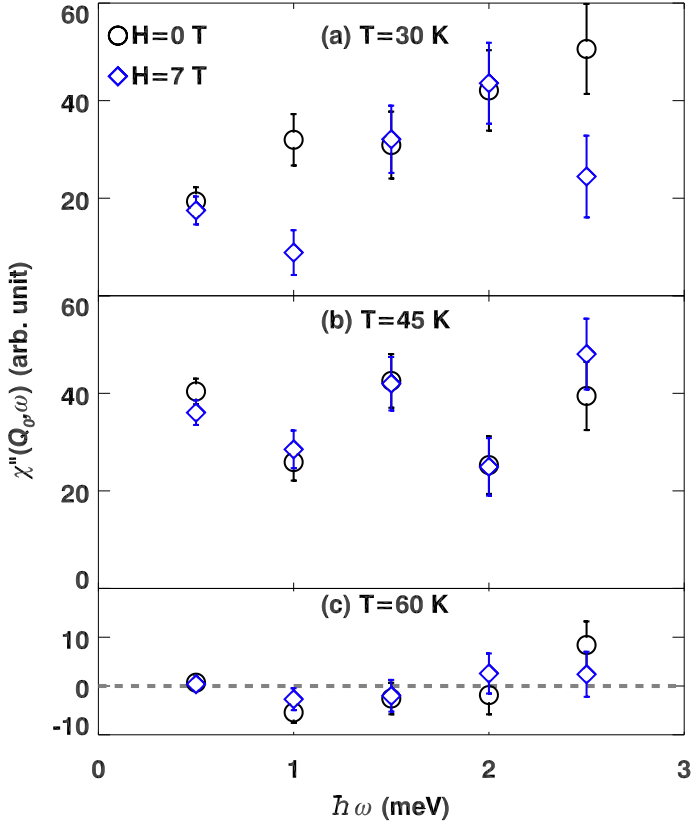}
\caption{(Color online) $\chi^{\prime\prime}({\bf Q}_0,\omega)$ with
${\bf Q}_{0}=(0.615,0.5,0)$ in zero field (circles) and 7~T field
(diamonds) at 30, 45, and 60~K converted from the peak intensity, as
discussed in the text. Error bars represent $1\pm\sigma$
uncertainties determined assuming Poisson statistics and those
smaller than the symbols are absent.} \label{fig:4}
\end{figure}

\begin{figure}[ht]
\includegraphics[width=\linewidth]{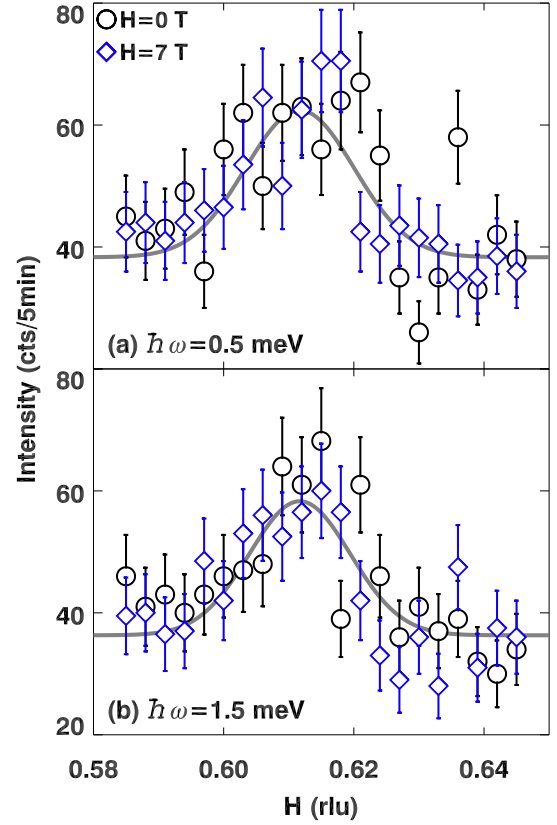}
\caption{(Color online) Scan profiles along ${\bf Q}=(H,0.5,0)$ at
30~K, with energies of $\hbar\omega=0.5$ and 1.5~meV, in zero field
(circles) and 7~T field (diamonds). Lines through the data are
guides for the eyes. Error bars represent the square root of the
counts.} \label{fig:5}
\end{figure}

Next we examine the field effect in finer detail by looking at
selected $(0.615+h,0.5,0)$ scans at 5 and 45~K (see
Fig.~\ref{fig:2}). At both 5 and 45~K, there are well defined peaks
at (0.615,0.5,0), well above the background, as represented by the
55~K data, although the peak at 45~K is much broader and the
intensity is weaker. At 5~K, where we have already seen that the
enhancement is relatively weak compared to that near 45~K,
zero-field and 7~T data are almost identical when measured with a
counting time of 1~min per point. At 45~K, the difference in
intensity is quite apparent---the enhancement is $\sim$20\%---while
the peak width shows little change.

We have applied different fields from 0 to 7~T at various
temperatures to check the field and temperature dependences of the
peak intensity; the results are shown in Fig.~\ref{fig:3}. It is
clear that with increasing magnetic field, the peak intensity
increases but only by a small amount.

We performed inelastic neutron-scattering measurements to study the
low-energy spin excitations. We scanned energy from 0.5 to 2.5~meV
at ${\bf Q}_0=(0.615,0.5,0)$ to look at the peak intensity's energy
dependence in fields of 0 and 7~T at various temperatures. The
intensity has been converted to the imaginary part of the dynamical
susceptibility $\chi^{\prime\prime}$ using
\begin{equation}\label{1}
    {\chi^{\prime\prime}({\bf Q}_{0},\omega)}=I({\bf Q}_0,\omega)(1-e^{{-\hbar\omega}/{k_{B}T}}),
\end{equation}
where $\omega$ is 2$\pi$ times frequency, $I({\bf Q}_0,\omega)$ is
the peak intensity, $\hbar$ is the Planck constant divided by
2$\pi$, $k_{B}$ is the Boltzmann constant, and $T$ is temperature.
The converted $\chi^{\prime\prime}({\bf Q}_{0},\omega)$ is plotted
in Fig.~\ref{fig:4}. At 60~K, $\chi^{\prime\prime}$ is negligible
(at the level of sensitivity in this experiment), and at 45~K, the
inelastic signal remains almost constant in the energy range from
0.5 to 2.5~meV. At 30~K, there seems to be a small gap at low
energy. These results agree well with those in
Ref.~\onlinecite{johnnew}, where it is shown that a gap opens at low
temperature in this La$_{1.875}$Ba$_{0.125}$CuO$_4$ crystal. After
applying a 7~T magnetic field, the inelastic signals do not seem to
be affected, as evidenced from $\chi^{\prime\prime}({\bf
Q}_{0},\omega)$.

The field effect is also absent in the {\bf Q} scans.
Constant-energy scans with $\hbar\omega=0.5$ and 1.5 meV along
$(0.615+h,0.5,0)$ in zero field and 7 T field for 30~K are plotted
in Fig.~\ref{fig:5}. These {\bf Q} scans are not distinguishable,
and no magnetic field impact on the gap is observable here. Since the spin
gap associated with superconductivity is rather sensitive to
magnetic field, the lack of field dependence seems to rule out a
connection between the spin gap and superconductivity. Most likely,
the gap is due to spin-orbit or exchange-anisotropy effects;
however, even a conventional spin-wave gap should be reduced by an
applied field due to Zeeman splitting of the spin-wave energies.
Clearly, much better counting statistics would be needed in order to
detect a finite change due to the field.

There is a sum rule for scattering from spin-spin correlations, and
hence one might ask whether the field induced enhancement of the
elastic peak should result in an observable decrease in the
inelastic magnetic scattering. Applying a 7-T field at low
temperature causes an increase in the elastic magnetic signal of
approximately 200 counts per 5 min of counting.  The measured energy
half-width of the elastic peak is 0.06 meV; thus, if this were
compensated by a decrease in inelastic scattering spread over an
energy range of 1 meV, we would expect to see a signal decrease of
about 12 counts per 5 min.  Looking at Fig.~\ref{fig:5}, such a
change would be big enough to be detectable.   One possible reason
that such an effect is not seen could be that the decrease in
scattering is spread over a significantly larger energy range, in
which case the effect would be in the noise.  Another possibility is
that the elastic enhancements come at the expense of spin degrees of
freedom associated with 2D superconducting correlations, as the
superconductivity is significantly depressed by the magnetic
field.~\cite{prl67001,johnnew}

To summarize, we have demonstrated that a {\it c}-axis magnetic
field shows its impact on the spin-stripe order by causing a slight
enhancement of the spin-order peak intensity, with no influence on
the peak width.  The biggest field effect on the intensity is near
the onset of spin order. Analysis of the peak width in zero field
reveals that the correlation length of the spin order along the
stripes is greater than that perpendicular to them. Finally, we have
seen a small spin gap with no significant magnetic field dependence.

The work at Brookhaven National Laboratory was supported by the U.S.
Department of Energy under Contract No.~DE-AC02-98CH10886. This work
utilized facilities supported in part by the National Science
Foundation under Agreement No. DMR-0454672.


\end{document}